\documentclass[aps,prl,twocolumn,showpacs,floatfix]{revtex4} 
\usepackage{amsmath,amssymb,isolatin1,graphicx,times}

\newcommand{\be}{\begin{equation}}
\newcommand{\ee}{\end{equation}}
\newcommand{\bea}{\begin{eqnarray}}
\newcommand{\eea}{\end{eqnarray}}
\newcommand{\bw}{\begin{widetext}}
\newcommand{\ew}{\end{widetext}}
\newcommand{\pavg}{\Pi_M(t)}
\newcommand{\kommentar}[1]{}
\newcommand{\rydn}{\textsf{n}}
\begin{document}
 
\title{Survival Probabilities in Coherent Exciton Transfer with Trapping
}
\author{Oliver M{\"u}lken}
\author{Alexander Blumen}
\author{Thomas Amthor}
\author{Christian Giese}
\author{Markus Reetz-Lamour}
\author{Matthias Weidem{\"u}ller}
\affiliation{
Physikalisches Institut, Universit\"at Freiburg,
Hermann-Herder-Stra{\ss}e 3, 79104 Freiburg i.Br., Germany}
 
\date{\today} 
\begin{abstract}
In the quest for signatures of coherent transport we consider exciton trapping in the continuous-time quantum walk framework. The survival probability displays different decay domains, related to distinct regions of the spectrum of the Hamiltonian. For linear systems and at intermediate times the decay obeys a power-law, in contrast to the corresponding exponential decay found in incoherent continuous-time random walk situations. To differentiate between the coherent and incoherent mechanisms, we present an experimental protocol based on a frozen Rydberg gas structured by optical dipole traps.
\end{abstract}
\maketitle


Recent years have seen an upsurge of interest in coherent energy transfer,
given the experimental advances in manipulating and controlling
quantum mechanical systems.  From the theoretical side, such
investigations are of long standing; see, e.g., \cite{Ziman}.  Here,
tight-binding models, which model coherent exciton transfer, are closely
related to the 
quantum walks (QW). As their classical random
walk (RW) counterpart, QW appear in two variants: discrete-time QW
\cite{aharonov1993} and continuous-time QW (CTQW) \cite{farhi1998}. 
Experimental implementations have
only recently been proposed for both QW variants, based, e.g., on microwave
cavities \cite{sanders2003}, ground state atoms \cite{duer2002} or Rydberg
atoms \cite{cote2006} in optical lattices, or the orbital angular momentum of
photons \cite{zhang2007}.

An appropriate means to monitor transport is to follow the decay of the
excitation due to trapping. The long time decay of chains with traps is a
well studied problem for classical systems
\cite{klafter1980,grassberger1982}: for an ensemble of chains of different
length with traps at both ends the averaged exciton survival probability
has a stretched exponential form $\exp(-bt^\lambda)$, with $\lambda=1/3$
(see, e.g., \cite{grassberger1982}). In contrast, quantum mechanical
tight-binding models lead to $\lambda=1/4$ \cite{excitontrap,parris1989b}.
However, up to now only little is known about the decay of the quantum
mechanical survival probability at experimentally relevant intermediate
times. 

Here we evaluate and compare the intermediate-time decays due to trapping for both RW and 
QW situations by employing the similarity of the CTRW and the CTQW formalisms.
Without traps, the coherent dynamics of excitons on a graph
of connected nodes is modeled by the CTQW, which is obtained by
identifying the Hamiltonian ${\bf H}_0$ of the system with the CTRW
transfer matrix ${\bf T}_0$, i.e., ${\bf H}_0 = - {\bf T}_0$; see e.g.\
\cite{farhi1998,mb2005a} (we will set $\hbar \equiv 1$ in the following).
For undirected graphs, ${\bf T}_0$ is related to the connectivity matrix
${\bf A}_0$ of the graph by ${\bf T}_0 = - {\bf A}_0$, where (for
simplicity) all transmission rates are taken to be equal. Thus, in the
following we take ${\bf H}_0 = {\bf A}_0$. The matrix ${\bf A}_0$ has as
non-diagonal elements $A^{(0)}_{k,j}$ the values $-1$ if nodes $k$ and $j$
of the graph are connected by a bond and $0$ otherwise. The diagonal
elements $A^{(0)}_{j,j}$ of ${\bf A}_0$ equal the number of bonds $f_j$
which exit from node $j$.  By fixing the coupling strength between two
connected nodes to $|H^{(0)}_{k,j}|=1$, the time scale is given in units
of $[\hbar/H^{(0)}_{k,j}]$. For the Rydberg gases considered in the
following, the coupling strength is roughly
$H^{(0)}_{k,j}/\hbar\gtrsim1$~MHz, i.e., the time unit for transfer
between two nodes is of the order of a few hundred nanoseconds.

The states $|j\rangle$ associated with excitons localized at the
nodes $j$ ($j=1,\dots,N$) form a complete, orthonormal basis set (COBS) of
the whole accessible Hilbert space, i.e., $\langle k | j \rangle =
\delta_{kj}$ and $\sum_k |k~\rangle\langle~k| = {\bf 1}$.  In general, the
time evolution of a state $|j\rangle$ starting at time $t_0=0$ is given by
$| j;t \rangle = \exp(-i {\bf H}_0 t)|j\rangle$; hence the transition
amplitudes and the probabilities read $\alpha_{kj}(t) \equiv \langle k |
\exp(-i {\bf H}_0 t) | j \rangle$ and $\pi_{kj}(t) \equiv \left|
\alpha_{kj}(t) \right|^2$, respectively. In the corresponding classical
CTRW case the transition probabilities follow from a master equation as
$p_{kj}(t) = \langle k | \exp({\bf T}_0 t) | j \rangle$
\cite{farhi1998,mb2005a}.


Consider now that out of the $N$ nodes $M$ are traps with $M\leq~N$; we denote
them by $m$, so that $m\in{\cal M}$, with ${\cal M}\subset\{1,\dots,N\}$.
We incorporate trapping into the CTQW formalism phenomenologically by
following an approach based on time dependent perturbation theory
\cite{excitontrap, parris1989b,Sakurai}. The new Hamiltonian is ${\bf H} =
{\bf H}_0 + i{\bf \Gamma}$, where the trapping operator $i{\bf \Gamma}$ has
at the trap nodes $m$ purely imaginary diagonal elements $i\Gamma_{mm}$,
which we assume to be equal for all $m$ ($\Gamma_{mm}\equiv\Gamma >0$), and
is zero otherwise. As a result, ${\bf H}$ is non-hermitian and has $N$
complex eigenvalues, $E_l = \epsilon_l - i\gamma_l$ ($l=1,\dots,N$). In
general, ${\bf H}$ has $N$ left and $N$ right eigenstates $|\Phi_l\rangle$
and $\langle\tilde\Phi_l|$, respectively. For most physically interesting
cases the eigenstates can be taken as biorthonormal, $\langle
\tilde\Phi_l | \Phi_{l'} \rangle = \delta_{ll'}$, and complete,
$\sum_{l=1}^N |\Phi_l\rangle\langle\tilde\Phi_l|={\bf 1}$; see, e.g.,
Ref.~\cite{sternheim1972}. Moreover, we have $\langle
k|\Phi_l\rangle^*=\langle\tilde\Phi_l|k\rangle$.  Thus, the transition
amplitudes can be calculated as $\alpha_{kj}(t) = \sum_l \exp[-\gamma_lt]
\exp[-i\epsilon_lt]\langle k | \Phi_l \rangle \langle \tilde\Phi_l | j
\rangle$; here the imaginary parts $\gamma_l$ of $E_l$ determine the
temporal decay of $\pi_{kj}(t) = \left| \alpha_{kj}(t) \right|^2$.


In an ideal experiment one would excite exactly one node, say
$j\not\in{\cal M}$, and read out the outcome $\pi_{kj}(t)$, i.e., the
probability to be at node $k\not\in{\cal M}$ at time $t$.  However, it is easier to
keep track of the total outcome at all nodes $k\not\in{\cal M}$, namely,
$\sum_{k\not\in{\cal M}} \pi_{kj}(t)$. Since the states $|k\rangle$ form a
COBS we have $\sum_{k\not\in{\cal M}} | k
\rangle\langle k | = {\bf 1} - \sum_{m\in{\cal M}} | m \rangle\langle m|$,
which leads to: 
\bea
\sum_{k\not\in{\cal M}} \pi_{kj}(t) 
&=& \sum_{l=1}^N  e^{-2\gamma_lt} \langle j | \Phi_l
\rangle\langle \tilde\Phi_l | j \rangle 
-\sum_{l,l'=1}^N e^{-i(E_l-E_{l'}^*)t} 
\nonumber \\
&& \times \sum_{m\in{\cal M}} \langle j |
\Phi_{l'} \rangle \langle \tilde\Phi_{l'} | m \rangle \langle m| \Phi_l
\rangle \langle \tilde\Phi_l | j \rangle.
\label{pi_avg_k}
\eea

By
averaging over all $j\not\in{\cal M}$, the mean survival probability 
$\pavg \equiv \frac{1}{N-M} \sum_{j\not\in{\cal M}} \sum_{k\not\in{\cal
M}} \pi_{kj}(t)$
is
given by 
\bea
&& \pavg 
= \frac{1}{N-M} \sum_{l=1}^N e^{-2\gamma_lt} \Big[ 1 - 2 \sum_{m\in{\cal
M}} \langle \tilde\Phi_l | m \rangle \langle m | \Phi_l \rangle \Big] 
\nonumber \\
&&+ \frac{1}{N-M} \sum_{l,l'=1}^N e^{-i(E_l-E_{l'}^*)t} \Big[
\sum_{m\in{\cal M}} \langle \tilde\Phi_{l'} | m \rangle \langle m| \Phi_l
\rangle \Big]^2.
\label{pi_avg}
\eea

For CTRW we include trapping in a formally similar fashion as for the
CTQW. Here, however, the classical transfer matrix ${\bf T}_0$ is
modified by the trapping matrix ${\bf \Gamma}$, such that the new
transfer matrix is ${\bf T} = {\bf T}_0 - {\bf \Gamma}$,
\cite{lakatos1972}. For a single linear system with traps at each end, the
mean survival probability $P_M(t) \equiv \frac{1}{N-M} \sum_{j\not\in{\cal
M}} \sum_{k\not\in{\cal M}} p_{kj}(t)$ decays exponentially at
intermediate and at long times \cite{lakatos1972}. As we proceed to show,
the decays of $\pavg$ and $P_M(t)$ are very different, thus allowing to
distinguish experimentally whether the exciton transfer is coherent or
not.

For long $t$ and small $M/N$, Eq.~(\ref{pi_avg}) simplifies considerably:
At long $t$ the oscillating term on the right hand side drops out and for
small $M/N$ we have $2 \sum_{m\in{\cal M}} \langle \tilde\Phi_l | m
\rangle\langle m | \Phi_l \rangle \ll~1$. Thus, $\pavg$ is mainly a sum of
exponentially decaying terms:
\be
\pavg \approx \frac{1}{N-M} \sum_{l=1}^N \exp[-2\gamma_lt].
\label{pi_avg2}
\ee

Asymptotically, Eq.~(\ref{pi_avg2}) is dominated by the $\gamma_l$ values
closest to zero. If the smallest one, $\gamma_{\rm min}$, is well
separated from the other values, one is led for 
$t \gg 1/\gamma_{\rm min}$ to the exponential decay found in earlier works,
$\pavg = \exp( - 2 \gamma_{\rm min} t)$ \cite{parris1989b}.

Such long times are not of much experimental relevance (see also below),
since most measurements highlight shorter times, in which many $\gamma_l$
contribute. In the corresponding energy range the $\gamma_l$ often scale,
as we show in the following, so that in a large $l$ range $\gamma_l \sim a
l^\mu$. The prefactor $a$ depends only on $\Gamma$ and $N$
\cite{parris1989b}. For densely distributed $\gamma_l$ and at intermediate
times one has, from Eq.~(\ref{pi_avg2}),
\be
\pavg \approx \int dx \ e^{- 2atx^\mu}
= \int dy \ \frac{e^{-y^\mu}}{(2at)^{1/\mu}} \sim t^{-1/\mu}.
\label{pi_avg_pl}
\ee


The envisaged experimental setup consists of clouds of ultra-cold Rydberg
atoms assembled in a chain over which an exciton migrates; the trapping of
the exciton occurs at the ends of the chain. The dipolar interactions
between Rydberg atoms depend on the mutual distance $R$ between the nodes
as $R^{-3}$. Now, CTRW over a chain of regularly arranged sites lead both
for nearest-neighbor steps and for step distributions depending on $R$ as
$R^{-\gamma}$, with $\gamma>3$, to a standard diffusive behavior and,
therefore, belong to the same universality class, see e.g.\ \cite{Weiss}.
The reason is that in one dimension for $\gamma>3$ the first two moments,
$\langle R \rangle$ and $\langle R^2 \rangle$, are finite. Thus, although
the {\sl quantitative} results will differ, the {\sl qualitative} behavior
is similar. Hence, we focus on a nearest-neighbor tight-binding model and
consider a chain of length $N$ with two traps ($M=2$) located at its ends
($m=1$ and $m=N$), 
\footnote{All numerical results were obtained by using FORTRAN's LAPACK
routines for diagonalizing non-hermitian matrices.}. 
The CTQW Hamiltonian thus reads 
\bea
{\bf H} &=& \sum_{n=1}^{N} \Big(2 | n \rangle \langle n| - | n-1 \rangle
\langle n | - | n+1 \rangle \langle n |\Big) \nonumber \\
&& + i\Gamma \sum_{m=1,N} | m
\rangle \langle m |.
\eea

\begin{figure}[htb]
\centerline{\includegraphics[clip=,width=\columnwidth]{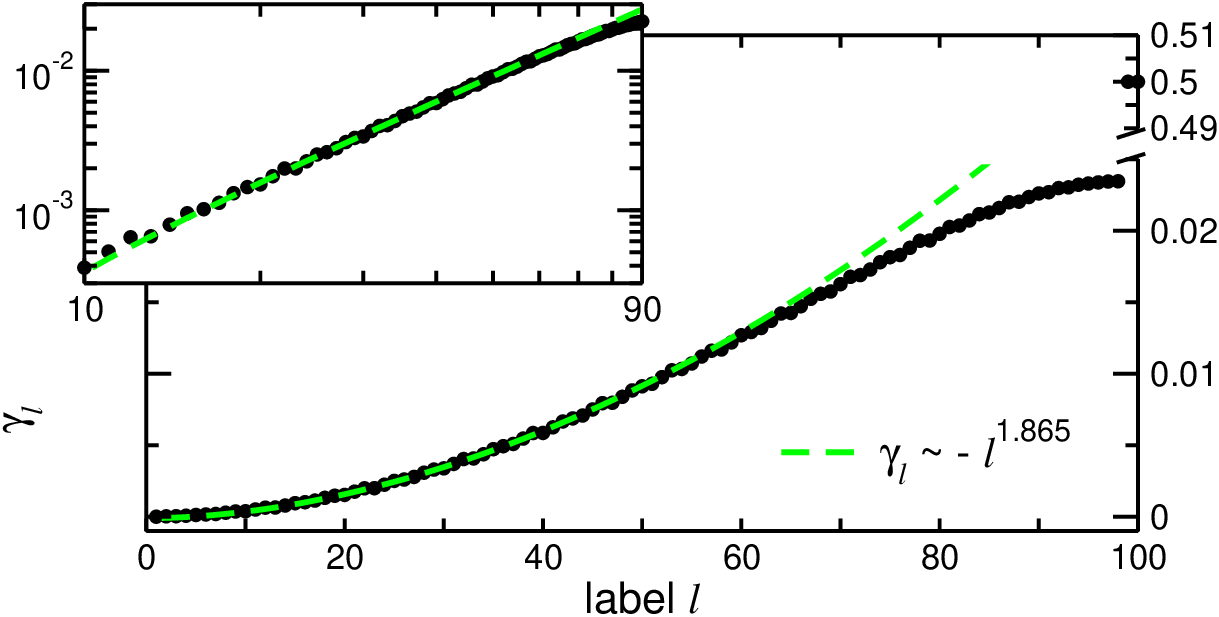}}
\caption{(Color online) Imaginary parts $\gamma_l$ (dots) in ascending
order for $N=100$ and $\Gamma=1$. Note the shortened $y$ axis.
The inset shows $\gamma_l$ in log-log scale for $l=10,\dots,90$.}
\label{evals_imag_n100}
\end{figure}

In Fig.~\ref{evals_imag_n100} we show the spectrum of $\gamma_l$ for
$N=100$ and $\Gamma=1$; the double logarithmic plot (see inset)
demonstrates that scaling holds for $10\leq l \leq 60$, with an exponent
of about $\mu=1.865$. In this domain $\gamma_l\in[0.0012,0.012]$, which
translates to experimentally accessible coherence times of
about $10-100\mu$s. For comparison the smallest decay rate is
$\gamma_{\rm min} = 7.94\times10^{-6}$, which corresponds to experimentally
unrealistic coherence times of the order of tenths of
seconds. 

\begin{figure}[htb]
\centerline{\includegraphics[clip=,width=\columnwidth]{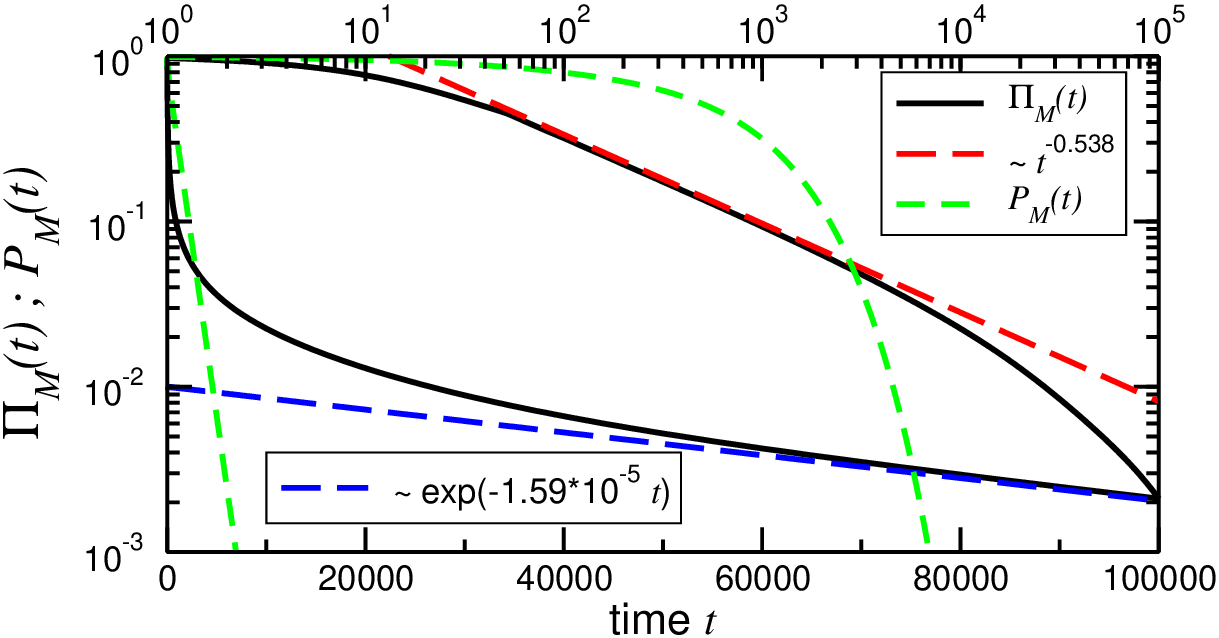}}
\caption{(Color online) Temporal decay of $\pavg$ (solid black lines) and
$P_M(t)$ (short dashed green lines) for $N=100$ and $\Gamma=1$ in double
logarithmic scales (upper three curves) and in logarithmic scales (lower
three curves). Indicated are the fits to $\pavg$ (long dashed lines) in the
intermediate (upper red) and the long (lower blue) time regime.}
\label{decay_n100}
\end{figure}

The corresponding transfer matrix 
of the classical CTRW 
reads 
\bea
{\bf T} &=& - \sum_{n=1}^{N} \Big(2 | n \rangle \langle n| - | n-1 \rangle
\langle n | - | n+1 \rangle \langle n |\Big) \nonumber \\
&& - \Gamma \sum_{m=1,N}
| m \rangle \langle m |.
\eea
In Fig.~\ref{decay_n100} we compare the classical $P_M(t)$ to the quantum
mechanical survival probability $\pavg$ for a linear system with $N=100$
and $\Gamma=1$.  Evidently, $P_M(t)$ and $\pavg$ differ strongly: the
$P_M(t)$ decay established for CTRW is practically exponential. $\pavg$,
on the other hand, shows two regimes: a power-law decay at intermediate
times (upper red curve) and an exponential decay
(lower blue curve) at very long times. 

\begin{figure}[htb]
\centerline{\includegraphics[clip=,width=\columnwidth]{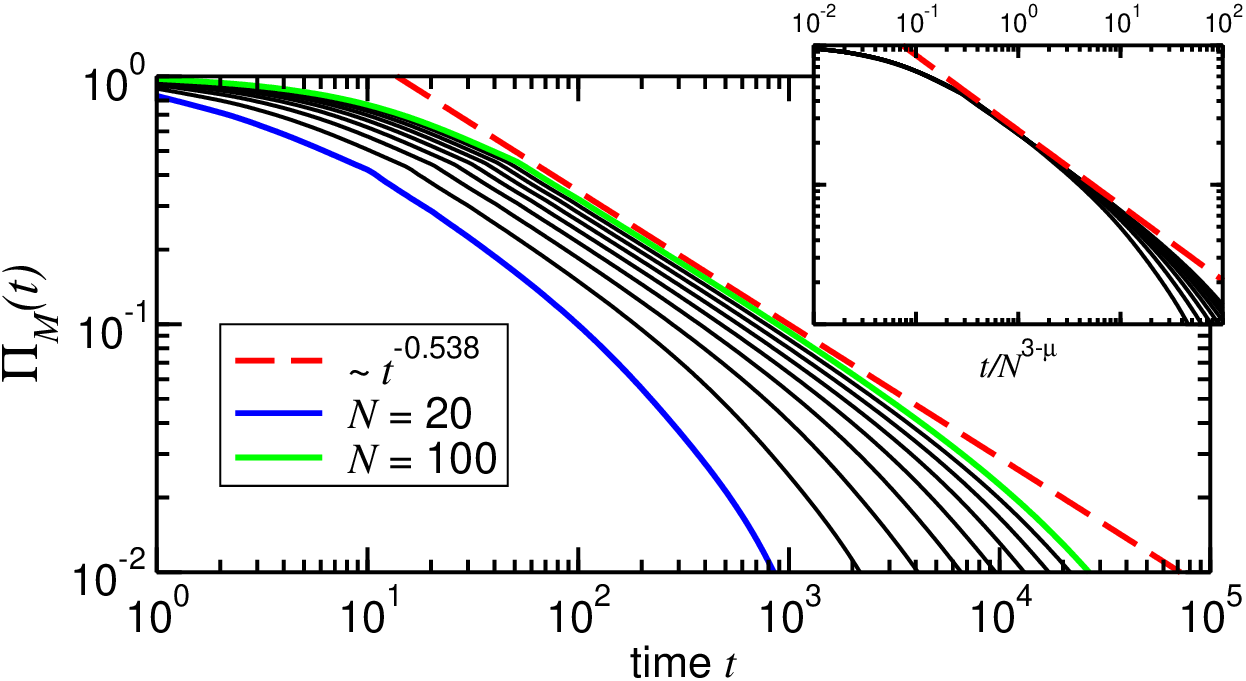}}
\caption{(Color online) $N$-dependence of  $\pavg$ for $\Gamma=1$; $N$
increases in steps of $10$ from $20$ (blue line) to $100$ (green line). The
inset shows $\pavg$ versus the rescaled time $t/N^{3-\mu}$.}
\label{avg_all_n}
\end{figure}

We now turn to the parameter dependences of $\pavg$.  Figure
\ref{avg_all_n} displays the dependence of $\pavg$ on $N$. We note that
the scaling regime, where $\pavg\sim t^{-1/\mu}$, gets larger with
increasing $N$. The cross-over to this scaling region from the domain of
short times occurs around $t\approx N/2$.  For larger $N$ and in the
intermediate time domain $\pavg$ scales nicely with $N$. In this case, the
power-law approximation [Eq.~(\ref{pi_avg_pl})] holds and by rescaling $l$
to $l/N$ we get from Eq.~(\ref{pi_avg2}) that
\be
\pavg \sim \sum_l e^{-2N^{-3}l^\mu t} 
= \sum_l \exp\Big[-2(l/N)^\mu
N^{-(3-\mu)}t\Big],
\label{tau_approx}
\ee
where we used that $a\sim N^{-3}$ for a linear system \cite{parris1989b}.
Thus when rescaling $l$ to $l/N$, time has to be rescaled by the factor
$N^{-(3-\mu)}$. Indeed, all curves where a power-law behavior can be
justified fall on a master curve; see the inset in Fig.~\ref{avg_all_n}.

\begin{figure}[htb]
\centerline{\includegraphics[clip=,width=\columnwidth]{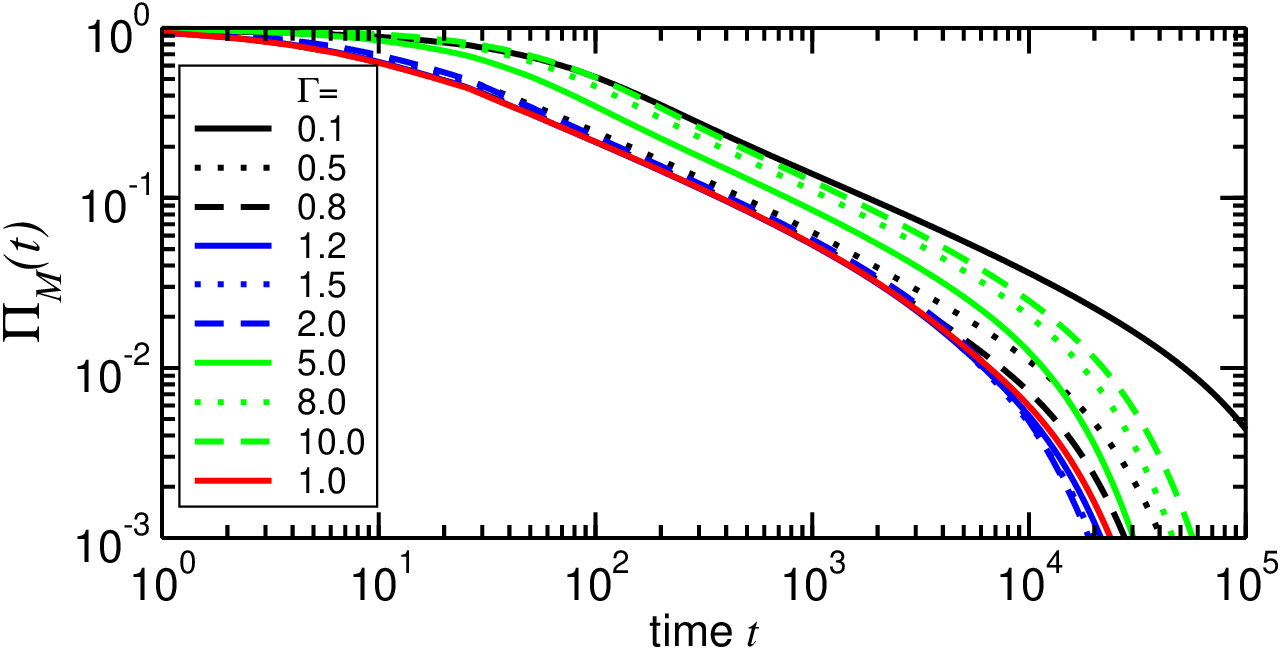}}
\caption{(Color online) $\Gamma$-dependence of $\pavg$ for intermediate
$t$ and $N=50$.}
\label{avg_all_g}
\end{figure}

The temporal decay does not only depend on $N$ but also on $\Gamma$.
Figure \ref{avg_all_g} shows $\pavg$ for $N=50$ and different $\Gamma$.
For values $\Gamma\gg1$ (green lines) and $\Gamma\ll1$ (black lines) the
curves shift to longer times. Values of $\Gamma$ close to $1$ (blue lines)
lead to the quickest decay. Note that these values are of the same order
as the coupling strength between the non trap nodes, $H_{j,j\pm1}=-1$. 

An experimental implementation of the described system has to meet several
criteria. A single node 
must represent a
well-defined two-level system to ensure coherent energy transfer while at
the same time a mechanism is needed to trap an exciton with a controllable
trapping efficiency. Furthermore, the chain must be static with
negligible motion and should allow for spatially selective excitation and
detection of the exciton. These demands rule out many possible candidates
for an experimental realization of CTQW. A frozen Rydberg gas [17] can
meet all of the above demands by combining the rich internal structure of
highly excited atoms whith the full quantum control over the external
degrees of freedom that is available in up-to-date experiments with
ultracold atoms. 
The
internal structure of Rydberg atoms provides both decoupled two-level
subsystems and tunable traps, while the pronounced Stark shift allows to
selectively address single sites in a chain when an electric field
gradient is applied. At the same time, experimentally accessible
temperatures below 1$\,\mu$K ensure that the thermal motion is negligible.

Our scheme starts from a cloud of laser-cooled ground state atoms prepared
in a chain of optical dipole traps~\cite{grimm00}. Each site represents
one node with distances between sites of 5 to 20\,$\mu$m. For an
experimentally achievable extension of 1\,mm this translates into
approximately 100 nodes.  All nodes are excited to Rydberg states
exploiting the dipole blockade mechanism to ensure a single Rydberg
excitation per node~\cite{lukin01} which avoids many-body
effects~\cite{anderson02}. A two-level system is realized by a transition
between energetically isolated states, i.e., by low-angular-momentum
states which exhibit a large quantum defect, e.g.,
$\rydn \textrm{S} \; \rightleftharpoons \; \rydn \textrm{P}\;$.
A number of experiments has revealed the coherent character of this
process~\cite{anderson02}. By contrast to low-$\ell$ states, states with
angular momenta $\ell\geq 3$ have no quantum defect and are degenerate.
This allows to construct an exciton trap with the transitions
$\rydn^\prime \textrm{D} \; \rightleftharpoons \; \rydn^{\prime\prime} \textrm{F}
\xrightarrow{\mathrm{rf}} \; \rydn^{\prime\prime} \ell (\ell\geq 3)$,
where the first transition is the dipole transition providing the coupling
to neighboring nodes
\footnote{In order to ensure the right coupling strength to neighboring
nodes both the energy difference and the transition dipole moments of
the processes 
$\rydn \textrm{S} \; \rightleftharpoons \; \rydn \textrm{P}\;$
and 
$\rydn^\prime \textrm{D} \; \rightleftharpoons \; \rydn^{\prime\prime}
\textrm{F}
\xrightarrow{\mathrm{rf}} \; \rydn^{\prime\prime} \ell (\ell\geq 3)$
must be the
same. For instance, in rubidium the pairs 71S/71P and 61D/60F fulfill this
condition at an  offset field of $\sim$70\,mV/cm with an energy difference
of $\Delta \mathrm{E}_{\mathrm{S/P}}=\Delta
\mathrm{E}_{\mathrm{D/F}}=h\,10.1\,$GHz and radial transition matrix
elements of 5200\,au and 4800\,au, respectively.}
while the second transition, driven by a radio-frequency (rf) field,
represents the trap and decouples this site from the energy transfer, as
the large degeneracy of the high-$\ell$ states ensures an efficient
suppression of the coupling back to the $\rydn^{\prime\prime}$F state
\footnote{Note that the rf frequency is detuned for any transitions in the
other nodes as those involve different atomic states.}.
By changing the strength of the driving rf field, the trapping efficiency
can be tuned. The population of the $\rydn^{\prime\prime}\ell$ state is directly
proportional to $1-\Pi_M(t)$ and can be determined by state selective
field ionization~\cite{gallagher94}. In an experiment the central nodes
would be prepared in the S state and the trap nodes in the D state. A
single S node is swapped to P through a microwave transition in an
electric field gradient which makes the resonance S$\rightarrow$P position
sensitive. This is equivalent to exciting a single exciton. The energy
transport is started by removing the field gradient making the transition
energy the same for all nodes.

There are two important decoherence mechanisms which are given by the
spontaneous decay of the involved Rydberg states and by the atomic motion.
Exemplarily, for the 71S and 61D states of rubidium and a distance of
20$\mu$m between nodes we calculate a transfer time of $\tau=$145\,ns
between two neighboring sites, radiative lifetimes including black-body
radiation of $\geq$100\,$\mu$s and residual thermal motion that leads to a
change of the interatomic distance of 1.4\,$\mu$m per 100\,$\mu$s at a
temperature of 1\,$\mu$K.  Another source of decoherence is the
interaction-induced motion~\cite{li05}. We can model this motion
quantitatively~\cite{amthor07} and calculate negligible changes of the
interatomic distances of less than 0.2\,$\mu$m per 100\,$\mu$s. This means
that both the chain and the elementary atomic system sustain coherence
over timescales on the order of several ten $\mu$s and longer.


In conclusion, we have identified different time domains in the CTQW
exciton decay in the presence of traps, domains which are directly
related to the complex spectrum of the system's Hamiltonian. The CTQW
average survival probability $\pavg$ for an exciton to stay inside a
linear system of $N$ nodes with traps at each end can clearly be
distinguished from its classical CTRW counterpart, $P_M(t)$. Finally, we
proposed an experimental test for coherence on the basis of ultra-cold
Rydberg atoms.


We gratefully acknowledge support from the Deutsche
For\-schungs\-ge\-mein\-schaft (DFG), the Ministry of Science,
Research and the Arts of Baden-W\"urttemberg (AZ: 24-7532.23-11-11/1) and
the Fonds der Chemischen Industrie.


\newpage

\begin{thebibliography}{19}
\expandafter\ifx\csname natexlab\endcsname\relax\def\natexlab#1{#1}\fi
\expandafter\ifx\csname bibnamefont\endcsname\relax
  \def\bibnamefont#1{#1}\fi
\expandafter\ifx\csname bibfnamefont\endcsname\relax
  \def\bibfnamefont#1{#1}\fi
\expandafter\ifx\csname citenamefont\endcsname\relax
  \def\citenamefont#1{#1}\fi
\expandafter\ifx\csname url\endcsname\relax
  \def\url#1{\texttt{#1}}\fi
\expandafter\ifx\csname urlprefix\endcsname\relax\def\urlprefix{URL }\fi
\providecommand{\bibinfo}[2]{#2}
\providecommand{\eprint}[2][]{\url{#2}}

\bibitem[{\citenamefont{Ziman}(1972)}]{Ziman}
\bibinfo{author}{\bibfnamefont{J.~M.} \bibnamefont{Ziman}},
  \emph{\bibinfo{title}{Principles of the Theory of Solids}}
  (\bibinfo{publisher}{Cambridge University Press, Cambridge, England},
  \bibinfo{year}{1972}).
  
\bibitem[{\citenamefont{Aharonov et~al.}(1993)\citenamefont{Aharonov,
  Davidovich, and Zagury}}]{aharonov1993}
\bibinfo{author}{\bibfnamefont{Y.}~\bibnamefont{Aharonov}},
  \bibinfo{author}{\bibfnamefont{L.}~\bibnamefont{Davidovich}},
  \bibnamefont{and}
\bibinfo{author}{\bibfnamefont{N.}~\bibnamefont{Zagury}},
  \bibinfo{journal}{Phys.\ Rev.\ A} \textbf{\bibinfo{volume}{48}},
  \bibinfo{pages}{1687} (\bibinfo{year}{1993});
%
\bibinfo{author}{\bibfnamefont{J.}~\bibnamefont{Kempe}},
  \bibinfo{journal}{Contemporary Physics} \textbf{\bibinfo{volume}{44}},
  \bibinfo{pages}{307} (\bibinfo{year}{2003}).
  
\bibitem[{\citenamefont{Farhi and Gutmann}(1998)}]{farhi1998}
\bibinfo{author}{\bibfnamefont{E.}~\bibnamefont{Farhi}} \bibnamefont{and}
  \bibinfo{author}{\bibfnamefont{S.}~\bibnamefont{Gutmann}},
  \bibinfo{journal}{Phys.\ Rev.\ A} \textbf{\bibinfo{volume}{58}},
  \bibinfo{pages}{915} (\bibinfo{year}{1998}).

\bibitem[{\citenamefont{Sanders et~al.}(2003)\citenamefont{Sanders,
Bartlett,
  Fregenna, and Knight}}]{sanders2003}
\bibinfo{author}{\bibfnamefont{B.~C.} \bibnamefont{Sanders}}
\bibinfo{author}{\bibfnamefont{{\sl et al.}}},
  \bibinfo{journal}{Phys.\ Rev.\ A} \textbf{\bibinfo{volume}{67}},
  \bibinfo{pages}{042305} (\bibinfo{year}{2003}).

\bibitem[{\citenamefont{D{\"u}r et~al.}(2002)\citenamefont{D{\"u}r,
  Raussendorf, Kendon, and Briegel}}]{duer2002}
\bibinfo{author}{\bibfnamefont{W.}~\bibnamefont{D{\"u}r}}
\bibinfo{author}{\bibfnamefont{{\sl et al.}}},
  \bibinfo{journal}{Phys.\ Rev.\ A}
  \textbf{\bibinfo{volume}{66}}, \bibinfo{pages}{052319}
  (\bibinfo{year}{2002}).

\bibitem[{\citenamefont{C{\^o}t{\'e} et~al.}(2006)\citenamefont{C{\^o}t{\'e},
  Russel, Eyler, and Gould}}]{cote2006}
\bibinfo{author}{\bibfnamefont{R.}~\bibnamefont{C{\^o}t{\'e}}}
\bibinfo{author}{\bibfnamefont{{\sl et al.}}},
  \bibinfo{journal}{New J.\ Phys.} \textbf{\bibinfo{volume}{8}},
  \bibinfo{pages}{156} (\bibinfo{year}{2006}).

\bibitem[{\citenamefont{Zhang et~al.}(2007)\citenamefont{Zhang, Ren, Zou,
Liu,
  Huang, and Guo}}]{zhang2007}
\bibinfo{author}{\bibfnamefont{P.}~\bibnamefont{Zhang}},
\bibinfo{author}{\bibfnamefont{{\sl et al.}}},
  \bibinfo{journal}{Phys.\ Rev.\ A} \textbf{\bibinfo{volume}{75}},
  \bibinfo{pages}{052310} (\bibinfo{year}{2007}).

\bibitem[{\citenamefont{Klafter and Silbey}(1980)}]{klafter1980}
\bibinfo{author}{\bibfnamefont{J.}~\bibnamefont{Klafter}}
\bibnamefont{and}
  \bibinfo{author}{\bibfnamefont{R.}~\bibnamefont{Silbey}},
  \bibinfo{journal}{J.\ Chem.\ Phys.} \textbf{\bibinfo{volume}{72}},
  \bibinfo{pages}{849} (\bibinfo{year}{1980});
%
  \bibinfo{journal}{{\sl ibid.}} 
  \textbf{\bibinfo{volume}{74}},
  \bibinfo{pages}{3510} (\bibinfo{year}{1981}).

\bibitem[{\citenamefont{Grassberger and Procaccia}(1982)}]{grassberger1982}
\bibinfo{author}{\bibfnamefont{P.}~\bibnamefont{Grassberger}}
\bibnamefont{and}
  \bibinfo{author}{\bibfnamefont{I.}~\bibnamefont{Procaccia}},
  \bibinfo{journal}{J.\ Chem.\ Phys.}
  \textbf{\bibinfo{volume}{77}},
  \bibinfo{pages}{6281} (\bibinfo{year}{1982});
%
\bibinfo{author}{\bibfnamefont{R.~F.} \bibnamefont{Kayser}}
\bibnamefont{and}
  \bibinfo{author}{\bibfnamefont{J.~B.} \bibnamefont{Hubbard}},
  \bibinfo{journal}{Phys.\ Rev.\ Lett.} \textbf{\bibinfo{volume}{51}},
  \bibinfo{pages}{79} (\bibinfo{year}{1983});
%
\bibinfo{author}{\bibfnamefont{G.}~\bibnamefont{Forgacs}},
  \bibinfo{author}{\bibfnamefont{D.}~\bibnamefont{Mukamel}},
\bibnamefont{and}
  \bibinfo{author}{\bibfnamefont{R.~A.} \bibnamefont{Pelcovits}},
  \bibinfo{journal}{Phys.\ Rev.\ B} \textbf{\bibinfo{volume}{30}},
  \bibinfo{pages}{205} (\bibinfo{year}{1984}).

\bibitem[{\citenamefont{Pearlstein}(1971)}]{excitontrap}
\bibinfo{author}{\bibfnamefont{R.~M.} \bibnamefont{Pearlstein}},
  \bibinfo{journal}{J.\ Chem.\ Phys.} \textbf{\bibinfo{volume}{56}},
  \bibinfo{pages}{2431} (\bibinfo{year}{1971});
%
\bibinfo{author}{\bibfnamefont{D.~L.} \bibnamefont{Huber}},
  \bibinfo{journal}{Phys.\ Rev.\ B} \textbf{\bibinfo{volume}{22}},
  \bibinfo{pages}{1714} (\bibinfo{year}{1980});
%
  \textbf{\bibinfo{volume}{45}},
  \bibinfo{pages}{8947} (\bibinfo{year}{1992});
%
\bibinfo{author}{\bibfnamefont{P.~E.} \bibnamefont{Parris}},
  \bibinfo{journal}{Phys.\ Rev.\ Lett.} \textbf{\bibinfo{volume}{62}},
  \bibinfo{pages}{1392} (\bibinfo{year}{1989}{\natexlab{a}});
%
\bibinfo{author}{\bibfnamefont{V.~A.} \bibnamefont{Malyshev}},
  \bibinfo{author}{\bibfnamefont{R.}~\bibnamefont{Rodr{\'i}guez}},
  \bibnamefont{and}
  \bibinfo{author}{\bibfnamefont{F.}~\bibnamefont{Dom{\'i}nguez-Adame}},
  \bibinfo{journal}{J.\ Lumin.} \textbf{\bibinfo{volume}{81}},
  \bibinfo{pages}{127} (\bibinfo{year}{1999}).

\bibitem[{\citenamefont{Parris}(1989{\natexlab{b}})}]{parris1989b}
\bibinfo{author}{\bibfnamefont{P.~E.} \bibnamefont{Parris}},
  \bibinfo{journal}{Phys.\ Rev.\ B} \textbf{\bibinfo{volume}{40}},
  \bibinfo{pages}{4928} (\bibinfo{year}{1989}{\natexlab{b}}).

\bibitem[{\citenamefont{M{\"u}lken and
Blumen}(2005{\natexlab{a}})}]{mb2005a}
\bibinfo{author}{\bibfnamefont{O.}~\bibnamefont{M{\"u}lken}}
\bibnamefont{and}
  \bibinfo{author}{\bibfnamefont{A.}~\bibnamefont{Blumen}},
  \bibinfo{journal}{Phys.\ Rev.\ E} \textbf{\bibinfo{volume}{71}},
  \bibinfo{pages}{016101} (\bibinfo{year}{2005}{\natexlab{a}});
%
  \bibinfo{journal}{{\sl ibid.}} \textbf{\bibinfo{volume}{71}},
  \bibinfo{pages}{036128} (\bibinfo{year}{2005});
%
  \bibinfo{journal}{{\sl ibid.}} \textbf{\bibinfo{volume}{73}},
  \bibinfo{pages}{066117} (\bibinfo{year}{2006}{\natexlab{b}}).

\bibitem[{\citenamefont{Sakurai}(1994)}]{Sakurai}
\bibinfo{author}{\bibfnamefont{J.}~\bibnamefont{Sakurai}},
  \emph{\bibinfo{title}{Modern Quantum Mechanics}}
  (\bibinfo{publisher}{Addison-Wesley, Redwood City, CA},
  \bibinfo{year}{1994}), \bibinfo{edition}{2nd} ed.,
\bibinfo{pages}{p.~342ff.}

\bibitem[{\citenamefont{Sternheim and Walker}(1972)}]{sternheim1972}
\bibinfo{author}{\bibfnamefont{M.~M.} \bibnamefont{Sternheim}}
  \bibnamefont{and} \bibinfo{author}{\bibfnamefont{J.~F.}
  \bibnamefont{Walker}}, \bibinfo{journal}{Phys.\ Rev.\ C}
  \textbf{\bibinfo{volume}{6}}, \bibinfo{pages}{114}
(\bibinfo{year}{1972}).

\bibitem[{\citenamefont{Lakatos-Lindenberg
  et~al.}(1971)\citenamefont{Lakatos-Lindenberg, Hemenger, and
  Pearlstein}}]{lakatos1972}
\bibinfo{author}{\bibfnamefont{K.}~\bibnamefont{Lakatos-Lindenberg}},
  \bibinfo{author}{\bibfnamefont{R.~P.} \bibnamefont{Hemenger}},
  \bibnamefont{and} \bibinfo{author}{\bibfnamefont{R.~M.}
  \bibnamefont{Pearlstein}}, \bibinfo{journal}{J.\ Chem.\ Phys.}
  \textbf{\bibinfo{volume}{56}}, \bibinfo{pages}{4852}
(\bibinfo{year}{1971});
%
\bibinfo{author}{\bibfnamefont{F.}~\bibnamefont{Dom{\'i}nguez-Adame}},
  \bibinfo{author}{\bibfnamefont{E.}~\bibnamefont{Maci{\'a}}},
  \bibnamefont{and}
  \bibinfo{author}{\bibfnamefont{A.}~\bibnamefont{S{\'a}nchez}},
  \bibinfo{journal}{Phys.\ Rev.\ B} \textbf{\bibinfo{volume}{51}},
  \bibinfo{pages}{878} (\bibinfo{year}{1995}).

\bibitem[{\citenamefont{Weiss}(1994)}]{Weiss}
\bibinfo{author}{\bibfnamefont{G.~H.} \bibnamefont{Weiss}},
  \emph{\bibinfo{title}{Aspects and Applications of the Random Walk}}
  (\bibinfo{publisher}{North-Holland, Amsterdam}, \bibinfo{year}{1994}),
\bibinfo{pages}{p.~78ff};
%
\bibinfo{author}{\bibfnamefont{J.}~\bibnamefont{Klafter}}
\bibinfo{author}{\bibfnamefont{{ \sl et al.}}},
  \bibinfo{journal}{Phys.\ Rev.\ A} \textbf{\bibinfo{volume}{35}},
  \bibinfo{pages}{3081} (\bibinfo{year}{1987}),
\bibinfo{pages}{Eqs.~(25)-(27)}. 
  
\bibitem[{\citenamefont{Anderson et~al.}(1998)\citenamefont{Anderson, Veale,
  and Gallagher}}]{anderson98}
\bibinfo{author}{\bibfnamefont{W.~R.} \bibnamefont{Anderson}},
  \bibinfo{author}{\bibfnamefont{J.~R.} \bibnamefont{Veale}}, \bibnamefont{and}
  \bibinfo{author}{\bibfnamefont{T.~F.} \bibnamefont{Gallagher}},
  \bibinfo{journal}{Phys. Rev. Lett.} \textbf{\bibinfo{volume}{80}},
  \bibinfo{pages}{249} (\bibinfo{year}{1998});
%
\bibinfo{author}{\bibfnamefont{I.}~\bibnamefont{Mourachko}}
\bibinfo{author}{\bibfnamefont{{\sl et al.}}},
  \bibinfo{journal}{{\sl ibid.}} \textbf{\bibinfo{volume}{80}},
  \bibinfo{pages}{253} (\bibinfo{year}{1998}).

\bibitem[{\citenamefont{Grimm et~al.}(2000)\citenamefont{Grimm,
  Weidem{\"u}ller, and Ovchinnikov}}]{grimm00}
\bibinfo{author}{\bibfnamefont{R.}~\bibnamefont{Grimm}},
  \bibinfo{author}{\bibfnamefont{M.}~\bibnamefont{Weidem{\"u}ller}},
  \bibnamefont{and} \bibinfo{author}{\bibfnamefont{Y.~B.}
  \bibnamefont{Ovchinnikov}}, \bibinfo{journal}{Adv. At. Mol. Opt. Phys.}
  \textbf{\bibinfo{volume}{42}}, \bibinfo{pages}{95} (\bibinfo{year}{2000}).

\bibitem[{\citenamefont{Lukin et~al.}(2001)\citenamefont{Lukin, Fleischhauer,
  C{\^o}t{\'e}, Duan, Jaksch, Cirac, and Zoller}}]{lukin01}
\bibinfo{author}{\bibfnamefont{M.~D.} \bibnamefont{Lukin}}
\bibinfo{author}{\bibfnamefont{{\sl et al.}}},
  \bibinfo{journal}{Phys. Rev. Lett.} \textbf{\bibinfo{volume}{87}},
  \bibinfo{pages}{037901} (\bibinfo{year}{2001}).

\bibitem[{\citenamefont{Anderson et~al.}(2002)\citenamefont{Anderson, Robinson,
  Martin, and Gallagher}}]{anderson02}
\bibinfo{author}{\bibfnamefont{W.~R.} \bibnamefont{Anderson}}
\bibinfo{author}{\bibfnamefont{{\sl et al.}}},
  \bibinfo{journal}{Phys. Rev. A}
  \textbf{\bibinfo{volume}{65}}, \bibinfo{pages}{063404}
  (\bibinfo{year}{2002});
%
\bibinfo{author}{\bibfnamefont{M.}~\bibnamefont{Mudrich}}
\bibinfo{author}{\bibfnamefont{{\sl et al.}}},
  \bibinfo{journal}{Phys. Rev. Lett.} \textbf{\bibinfo{volume}{95}},
  \bibinfo{pages}{233002} (\bibinfo{year}{2005});
%
\bibinfo{author}{\bibfnamefont{S.}~\bibnamefont{Westermann}}
\bibinfo{author}{\bibfnamefont{{\sl et al.}}},
  \bibinfo{journal}{Eur. Phys. J. D} \textbf{\bibinfo{volume}{40}},
  \bibinfo{pages}{37} (\bibinfo{year}{2006}).

\bibitem[{\citenamefont{Gallagher}(1994)}]{gallagher94}
\bibinfo{author}{\bibfnamefont{T.~F.} \bibnamefont{Gallagher}},
  \emph{\bibinfo{title}{{R}ydberg Atoms}} (\bibinfo{publisher}{Cambridge
  University Press}, \bibinfo{address}{Cambridge}, \bibinfo{year}{1994}).

\bibitem[{\citenamefont{Li et~al.}(2005)\citenamefont{Li, Tanner, and
  Gallagher}}]{li05}
\bibinfo{author}{\bibfnamefont{W.}~\bibnamefont{Li}},
  \bibinfo{author}{\bibfnamefont{P.~J.} \bibnamefont{Tanner}},
  \bibnamefont{and} \bibinfo{author}{\bibfnamefont{T.~F.}
  \bibnamefont{Gallagher}}, \bibinfo{journal}{Phys. Rev. Lett.}
  \textbf{\bibinfo{volume}{94}}, \bibinfo{pages}{173001}
  (\bibinfo{year}{2005}).

\bibitem[{\citenamefont{Amthor et~al.}(2007)\citenamefont{Amthor, Reetz-Lamour,
  Westermann, Denskat, and Weidem{\"u}ller}}]{amthor07}
\bibinfo{author}{\bibfnamefont{T.}~\bibnamefont{Amthor}}
\bibinfo{author}{\bibfnamefont{{\sl et al.}}},
  \bibinfo{journal}{Phys. Rev. Lett.} \textbf{\bibinfo{volume}{98}},
  \bibinfo{pages}{023004} (\bibinfo{year}{2007}).
\end{thebibliography}
\end{document}